\documentclass[prd,nofootinbib,amsmath,superscriptaddress,11pt]{revtex4}

\usepackage{graphicx}
\usepackage[normalem]{ulem}	
\usepackage[usenames]{color}    
\usepackage{wasysym}
\usepackage{amsmath}

\usepackage{epsfig}

\def\lsim{\raise0.3ex\hbox{$\;<$\kern-0.75em\raise-1.1ex
\hbox{$\sim\;$}}}
\def\gsim{\raise0.3ex\hbox{$\;>$\kern-0.75em\raise-1.1ex
\hbox{$\sim\;$}}}

\def\be{\begin{equation}}
\def\ee{\end{equation}}
\def\ba{\begin{eqnarray}}
\def\ea{\end{eqnarray}}

\begin{document}
\title{A natural SM-like 126 GeV  Higgs  via non-decoupling D-terms}
\author{Enrico~Bertuzzo}
\email{ebertuzzo@ifae.es}
\affiliation{IFAE, Universitat Aut\`onoma de Barcelona, 08193 Bellaterra, Barcelona, Spain}
\author{Claudia~Frugiuele}
\email{claudiaf@fnal.gov}
\affiliation{Fermilab, P.O. Box 500, Batavia, IL 60510, USA}


\begin{abstract} 

Accommodating  both  a $126$ GeV mass and Standard Model (SM) like couplings for the Higgs  has a fine tuning price in supersymmetric models. Examples are the MSSM, in which SM-like couplings are natural, but raising the Higgs mass up to 126 GeV requires a considerable tuning, or the NMSSM, in which the situation is reversed: the Higgs is naturally heavier, but being SM-like  requires some tuning. We show that  models with non-decoupling $D$-terms alleviate this tension - a 126 GeV SM-like Higgs comes out basically with no fine tuning cost. In addition, the analysis of the fine tuning of the extended gauge sector shows that naturalness requires the heavy gauge bosons to likely be within the LHC run II reach.

\end{abstract} 

\begin{flushright}{\small FERMILAB-PUB-14-507-T}\end{flushright}

\maketitle

\tableofcontents
\section{Introduction}
The naturalness problem of supersymmetric (SUSY) theories is a long standing one. Already after LEP-II data, accommodating the Higgs boson mass in the minimal supersymmetric standard model (MSSM) required large radiative corrections, with a tuning already below the $10\%$ level~\cite{Barbieri:2000gf, Dine:2007xi}. The problem has become more acute after the Higgs discovery, since a mass of $126$ GeV requires a tuning worse than $1$ part in $75$~\cite{Hall:2011aa}. There is no firm theorem stating that a theory with such a tuning have to be discarded; still, it may be seen as an indication to go beyond the MSSM. 

It is clear that in order to improve the fine tuning, the Higgs sector must be cleverly modified. Broadly speaking, this can be done in two ways: either increasing the Higgs boson mass at tree level (like in the NMSSM~\cite{Fayet:1974pd, Ellwanger:2009dp}, in the triplet extended MSSM~\cite{Espinosa:1991wt, Espinosa:1991gr} or in models with non decoupling $D$-terms, on which we will focus~\cite{Batra:2003nj}), or enlarging the particle content of the theory to arrange for additional ``stop-like'' loop contribution (as happens in $R$-symmetric models~\cite{Bertuzzo:2014bwa, Diessner:2014ksa}).
As a consequence, accommodating a $126$ GeV Higgs does not necessarily represent a challenge for naturalness in MSSM extensions. However, the LHC has introduced two new naturalness probes in the picture: sparticles direct searches and Higgs couplings measurements. Let us discuss them in turn. 

Direct searches are certainly very powerful tools, but are strongly dependent on the detailed topologies appearing in sparticle decay chains. For instance, the lower bounds on gluino and stop masses depend crucially on the lightest neutralino mass~\cite{Aad:2014kra}. Moreover, they can be completely modified if the $R$-parity requirement is dropped, or if an $R$-symmetry is imposed on the theory~\cite{Chatrchyan:2013gia, TheATLAScollaboration:2013xia}. $R$-symmetric models also change dramatically the bounds coming from rare flavor decays like $b\rightarrow s\gamma$~\cite{Kribs:2007ac}, which may otherwise put significant bounds on the sparticle spectrum~\cite{Katz:2014mba}.

On the contrary, constraints extracted from Higgs physics are more robust, and can be used to place almost model independent bounds on the sparticle masses. More precisely, under mild assumptions the Higgs-gluon-gluon coupling can be used to extract lower bounds on the stop masses~\cite{D'Agnolo:2012mj, Gupta:2012fy, Farina:2013ssa, Fan:2014txa}, while the tree level couplings to fermions and vectors can be used to extract informations on the spectrum of the remaining CP-even scalars. As we are going to see, heavy additional scalars do not  require an effective fine tuning price only for sufficiently large $\tan\beta$ . 
The immediate consequence is that in the MSSM a natural SM-like Higgs can be obtained, with the $126$ GeV mass setting the fine tuning of the model. In contrast, models like the NMSSM, which requires small $\tan\beta$ to increase the Higgs mass at tree level, may have problems in accommodating natural SM-like couplings, since having the other scalars significantly heavier than the $126$ GeV Higgs requires a considerable tuning. After run I, this fine tuning price is still small, compared to direct searches constraints, but run-II with $300$ fb$^{-1}$ will be able to constrain the tuning at the few percent level~\cite{Farina:2013fsa, Gherghetta:2014xea}. Precision Higgs physics is therefore a powerful way to test naturalness in the NMSSM, and it has been shown to be effective also in models of uncolored naturalness, both supersymmetric and not~\cite{Burdman:2014zta, Craig:2013fga}.

The purpose of this paper is to show that in models with non decoupling $D$-terms a tuning better than $20\%$ can accommodate both a $126$ GeV mass and no deviations in Higgs couplings even after run II of the LHC. Even future colliders like the ILC and TLEP will be able to probe the fine tuning only up to the $10\%$ level. This implies that Higgs precision physics will not be an effective probe of naturalness in this framework, leaving the probe of the natural parameter space to direct searches. Interestingly, as we are going to show, a low fine tuning requires the heavy gauge bosons to likely be in the LHC run II reach, adding a new naturalness probe to those already given by direct searches of squarks, gluinos and higgsinos.

Models with non decoupling $D$-terms have been studied in~\cite{Batra:2003nj, Batra:2004vc, Maloney:2004rc, Medina:2009ey, Cheung:2012zq, Craig:2012bs, Huo:2012tw, Athron:2009bs, Athron:2012sq, Athron:2013ipa, Bharucha:2013ela}, and in~\cite{Lodone:2010kt} the fine tuning was studied for a heavy Higgs boson.  In~\cite{Blum:2012ii, D'Agnolo:2012mj} the Higgs couplings deviations from the SM behavior were studied in the effective theory below the heavy vectors threshold, but with a different emphasis and without discussing fine tuning implications.

\section{Setting up the tools: fine tuning computation}\label{sec:tools}

In this section we give our general definition of fine tuning and make contact with the standard definitions~\cite{Barbieri:1987fn, Dimopoulos:1995mi, Kitano:2005wc}.
To this purpose we start considering the following potential:
\begin{equation}\label{eq:potential_2HDM_loop}
 V= m_u^2  \, |H_u|^2 + m_d^2 \, |H_d|^2 + B \, H_u \, H_d + h.c. + \lambda_{tree} \, \left(|H_u|^2 - |H_d|^2\right)^2 + \lambda_u \, |H_u|^4 + \lambda_{ud} |H_u H_d|^2\, ,
\end{equation}
which is a simplified form of the full Coleman Weinberg potential (see~\cite{Cassel:2009ps, Cassel:2010px} for early works where the minimization is done for the full Coleman Weinberg potential). In Eq.~(\ref{eq:potential_2HDM_loop}), $\lambda_{tree}$ indicates a tree level coupling (either the standard supersymmetric $D$-terms or the modified expression arising in non decoupling $D$-terms models, Sec.~\ref{sec:naturalness_bounds}), while $\lambda_u$ and $\lambda_{ud}$ parametrize possible additional tree or loop level corrections. For example, $\lambda_{ud}$ may correspond to the $F$-term quartic associated with the singlet in the NMSSM, while $\lambda_u$ may be a typical loop contribution from stops, or may arise when the Higgs couples to SUSY-breaking mediators for very low SUSY breaking scale. We stress that this approach of including the complete CW potential changes quantitatively the fine tuning measure with respect to the usual minimization at tree level. Since loop corrections may be numerically relevant, we believe their inclusion to be important in assessing the tuning of a model. 

If $\lambda_{tree}$ differs from the SUSY $D$-term contribution, $\lambda_{tree}^D = \frac{g^2 + g'^2}{8}$, it contributes together with $\lambda_u$ to an effective hard SUSY breaking in the low energy potential. We can estimate the contribution to the Higgs mass that diverge quadratically as
\begin{equation}
  \begin{array}{ccl}
 V &=& \frac{\Lambda^2}{32\pi^2} {\rm Str}M^2 + \dots \\
   &=&  \frac{N_p(\lambda_{tree} + \lambda_u) \Lambda^2}{32\pi^2} |H_u^0|^2 + \dots \, ,
 \end{array}
\end{equation}
where we assume the sum over the different contributions to be $N_p (\lambda_{tree}+\lambda_u) \simeq {\cal O}(1)$. From the associated tuning,
\begin{equation}\label{eq:estimate_tuning_quadratic}
 \Delta_{\Lambda^2} = \frac{\delta m^2_h}{m^2_h} \sim \frac{1}{32\pi^2} \frac{\Lambda^2}{m_h^2}\, ,
\end{equation}
we get that the theory is basically untuned, {\it i.e.} $\Delta_{\Lambda^2} <5$, for a cut off $\Lambda \lesssim 5$ TeV. In Section~\ref{sec:naturalness_bounds} we will show that this rough estimate agrees with the calculation done in the complete model. This strongly suggests that new physics leading to modified $D$-terms may naturally be in the LHC-13 reach, making the study of these models even more interesting.\\

Minimizing Eq.~(\ref{eq:potential_2HDM_loop}) we obtain
\begin{equation}\label{eq:general_minimum}
 v^2 \simeq  \frac{ s^2_\beta \, m^2_u -c^2_\beta \, m_d^2}{2 (\lambda_{tree} \, c_{2\beta} - \lambda_u s^4_\beta)}\, , ~~~~ 
 \frac{2B}{s_{2\beta}} \simeq \frac{ -2 \lambda_{tree} c_{2\beta} (m_u^2 + m_d^2) + 2 \lambda_u s_\beta^2 m_d^2 + \lambda_{ud} (m_d^2 c_\beta^2 - m_u^2 s_\beta^2 ) }{2 (\lambda_{tree} \, c_{2\beta} - \lambda_u s^4_\beta)} \, ,
\end{equation}
which can be used to compute the mass of the CP-odd state, $m^2_A = -2B/s_{2\beta}$, 
and the CP-even mass matrix in the vev basis $(h,H)$,
\begin{equation}\label{eq:general_masses}
 {\cal M}^2 =
 \begin{pmatrix}
  4 (\lambda_{tree} c^2_{2\beta} + \lambda_u s^4_\beta + \frac{1}{4}\lambda_{ud} s^2_{2\beta}) v^2 & \big( ( 4\lambda_{tree} -\lambda_{ud}) c_{2\beta} - 2 \lambda_u s_\beta^2 \big) v^2 s_{2\beta} \\
  \big( ( 4\lambda_{tree} -\lambda_{ud}) c_{2\beta} - 2 \lambda_u s_\beta^2 \big) v^2 s_{2\beta} & m_A^2 + (4\lambda_{tree} +\lambda_u -\lambda_{ud}) v^2 s_{2\beta}^2 
 \end{pmatrix}\, .
\end{equation}

Eq.~(\ref{eq:general_minimum}) can also be used to compute the sensitivity of the EW scale to the fundamental parameters $\xi_i$. Adopting the usual fine tuning measure~\cite{Barbieri:1987fn, Dimopoulos:1995mi},
\begin{equation}
 \Delta = {\rm max}_{\xi_i} \left| \frac{\delta \log v^2}{\delta \log \xi_i} \right| \, ,
\end{equation}
we get that variations of $m_u^2$ and $m_d^2$ lead to
\begin{equation}\label{eq:FT_general}
\begin{array}{ccl}
 {\displaystyle \Delta_{m_u^2}} &=& {\displaystyle \frac{m_u^2}{v^2} \frac{2 v^2 s_\beta^2 \left( m_A^2 + 2 v^2 c_\beta^2 (\lambda_{ud}-4 \lambda_{tree}) \right)}{ m_A^2 {\cal M}^2_{hh} + \left( \lambda_{ud}^2 -4 \lambda_{tree} (\lambda_u + \lambda_{ud})\right) v^4 s_{2\beta}^2 } \, ,} \\[0.3cm]
 {\displaystyle \Delta_{m_d^2}} &=& {\displaystyle \frac{m_d^2}{v^2} \frac{2 v^2 c_\beta^2 \left( m_A^2 + 2 v^2 s_\beta^2  (\lambda_{ud}-4 \lambda_{tree}) \right)}{ m_A^2 {\cal M}^2_{hh} + \left( \lambda_{ud}^2 -4 \lambda_{tree} (\lambda_u + \lambda_{ud}) \right) v^4 s_{2\beta}^2 } \, ,}
\end{array}
\end{equation}
which is the main result of this section. It can be used to study the tuning of any theory in which $H_u$ and $H_d$ are the only scalars remaining in the low energy theory, with an effective potential given by Eq.~(\ref{eq:potential_2HDM_loop}).

For large $\tan\beta$, the lightest scalar corresponds to $h$, with mass $m_h^2 \simeq {\cal M}_{hh}^2$. In this limit Eqs.~(\ref{eq:FT_general}) simplify to
\begin{equation}\label{eq:FT_general_largetb}
 \begin{array}{ccl}
  \displaystyle \Delta_{m_u^2} &\simeq & \displaystyle \frac{2 m_u^2}{m_h^2} \, ,\\[0.3cm]
  \displaystyle \Delta_{m_d^2} &\simeq & \displaystyle \frac{2 m_d^2}{m_h^2}\frac{m_A^2 - 2 v^2 (4 \lambda_{tree} + 2 \lambda_u - \lambda_{ud})}{m_A^2} \frac{1}{t_\beta^2} \, .\\
 \end{array}
\end{equation}
The first sensitivity may be used to compute the naturalness bounds on the Higgsino, stop and gluino masses, and for large $\tan\beta$ corresponds to the Kitano-Nomura measure~\cite{Kitano:2005wc}. Expanding Eq.~(\ref{eq:FT_general}) for small $v/m_A$ (a good approximation already for $m_A \gtrsim 250$ GeV), the computation of the tuning on the parameters $\{ \mu, m_{\tilde t}, M_3\}$ gives
\begin{equation}\label{}
 \begin{array}{ccl}
  {\displaystyle \mu} &\lesssim & {\displaystyle 140 \, {\rm GeV} \frac{1}{s_\beta}\left(\frac{{\cal M}^2_{hh}}{(126\, {\rm GeV})^2} \right)^{1/2} \left(\frac{\Delta}{5} \right)^{1/2} \, ,} \\
  {\displaystyle m_{\tilde t} } & \lesssim & {\displaystyle 600 \, {\rm GeV}  \left(\frac{{\cal M}^2_{hh}}{(126\, {\rm GeV})^2} \right)^{1/2} \left( \frac{3}{\log\frac{\Lambda}{\rm TeV}} \right)^{1/2} \left(\frac{\Delta}{5} \right)^{1/2} \, ,} \\
  {\displaystyle M_3 } & \lesssim & {\displaystyle 770 \, {\rm GeV}  \left(\frac{{\cal M}^2_{hh}}{(126\, {\rm GeV})^2} \right)^{1/2} \left( \frac{12}{\log\frac{\Lambda}{\rm TeV} \left( 1+ \log\frac{\Lambda}{\rm TeV} \right)} \right)^{1/2} \left(\frac{\Delta}{5} \right)^{1/2} \, ,}
 \end{array}
\end{equation}
where the parameters appearing on the left hand side are evaluated at the scale $\Lambda$ at which the RGE evolution starts. Notice that the bounds on $\mu$ and $M_3$ differ a factor $\sqrt{2}$ from those usually found in the literature because we compute the sensitivity with respect to $\mu$ and $M_3$ themselves, rather than $\mu^2$ and $M_3^2$.\\

The consequences of the second sensitivity have been less explored in the literature (see~\cite{Farina:2013fsa, Gherghetta:2014xea, Katz:2014mba} for three recent papers on the subject). Since $m_d^2$ roughly sets the $H$, $A$ and $H^\pm$ mass scale,~\footnote{Similar bounds can be obtained considering a variation of the $B$ parameter.} $\Delta_{m_d^2}$ measures the fine tuning on the EW scale due to the other scalars. For large $\tan{\beta}$, heavy scalars do not introduce a severe tuning, since the bound scales as $m_d^2/t_\beta^2$. On the contrary, for low or moderate $\tan{\beta}$ we expect the heavy scalars to be an important source of tuning.\\

The situation can thus be broadly summarized as follows: for small $\tan\beta$, in addition to Higgsinos, stops and gluino (entering respectively at tree, one and two loop level), also the tree level contribution due to $m_d^2$ may be subject to an important naturalness bound. At the phenomenological level, we know that stop and gluino searches are likely to be powerful enough to put relevant bounds on these masses (with specific bounds depending on the sparticle spectrum and to whether $R$-parity or an $R$-symmetry is imposed), while $\mu$ will likely be less constrained, due to the challenges in the Higgsino searches. $m_d^2$ will instead be a good probe of naturalness, since it controls the mixing of the lightest scalar with the heavier states and is thus going to be bounded by Higgs precision physics. This is relevant for instance in the NMSSM~\cite{Farina:2013fsa, Gherghetta:2014xea}, or in models in which the little hierarchy problem is solved by uncolored particles~\cite{Burdman:2014zta}.

For large $\tan\beta$, on the contrary, $m_d^2$ does not introduce any relevant tuning. In addition, in this limit the lightest scalar is already SM-like almost independently on the other scalar masses, Eq.~(\ref{eq:general_masses}), so that Higgs precision physics will hardly play any role as direct naturalness probe. We thus expect the usual direct stop and gluino searches to be the most powerful probes of naturalness in this regime. Nevertheless, with only mild assumptions, Higgs precision physics can be used to place important bounds on the loop stop contribution to the Higgs-gluon-gluon coupling (see~\cite{Farina:2013ssa, Fan:2014txa} and Sec.~\ref{sec:Higgs_couplings}).\\

Given the model dependence of the bounds from direct searches, in the following we will analyze only the possible tuning coming from Higgs coupling measurements. We stress however that this tuning is the minimum one for a given model. Once the complete framework is defined, bounds coming from direct searches have to be taken into account in assessing the overall fine tuning. Let us give some examples. In general, the bound most relevant for naturalness is the one on the gluino mass, once the stops are decoupled from the problem of the Higgs boson mass. Assuming $R$-parity conservation, the most constraining limit is $m_{\tilde g} \gtrsim 1400$ GeV~\cite{Aad:2014wea}. Taking into account the gluino mass running, $M_3(\mu)/g_s^2(\mu) \simeq {\rm const}$, we get $M_3 \gtrsim 1300$ GeV at a scale $\Lambda = 20$ TeV, {\it i.e.} $\Delta_{M_3} \gtrsim 15$. Assuming instead baryonic $R$-parity violation, the experimental bound gets relaxed to $m_{\tilde g} \gtrsim 800-900$ GeV at the TeV scale~\cite{CMS:2013bea, TheATLAScollaboration:2013xia}. Again at $\Lambda = 20$ TeV, we obtain $M_3 \gtrsim 700-800$ GeV, with a tuning $\Delta_{M_3} \gtrsim 5$.

\section{ An extended  gauge group as source of hard SUSY breaking}

As showed in the  previous section,  a natural UV completion that generates a hard SUSY breaking quartic coupling  should emerge at $ \Lambda \lesssim 5$ TeV, a scale  possibly  testable at LHC.
This is an important feature which deserves a complete study to make robust statements about the LHC phenomenology of a natural spectrum in this framework.
A quartic coupling $\lambda_{tree} \neq \frac{g^2+g'^2}{8}$ may be generated extending the SM gauge group and charging the Higgs fields under the new force.
The new gauge group must be broken below the SUSY breaking scale, to avoid the decoupling of the new contribution once the heavy gauge bosons are integrated out. 
These non decoupling $D$-terms are easily generated both in abelian extensions such as $SU(2)_L \times U(1)_Y \times U(1)_X$, as well as in non Abelian extensions such as $ SU(2)_A \times SU(2)_B \times U(1)_Y \rightarrow SU(2)_{L}\times U(1)_Y.$ Another well motivated possibility is offered by quiver groups in which the SM gauge group is doubled, so that both an extra $U(1) $ and an extra $SU(2) $ are present. 
Clearly, the hierarchy between the heavy gauge bosons and the soft SUSY breaking scale, necessary to generate non-decoupling D-terms, could turn into a new relevant source of tuning. 
In the present section we  quantify exactly this new possible tuning, and we show that the hierarchy does not need to be large to accommodate a $126$ GeV Higgs boson.
For simplicity we discuss the naturalness implications coming from the extended gauge sector, focusing on a particular UV completion, $ SU(2)_A \times SU(2)_B \times U(1)_Y \rightarrow SU(2)_{L}\times U(1)_Y$, which produces non decoupling $D$-terms. We then show how a $126$ GeV SM-like Higgs boson is a natural outcome in a large region of the parameter space.

\subsection{ Naturalness bounds from the extended gauge sector}\label{sec:naturalness_bounds}

We now analyze the simplest non abelian extension of the SM electroweak gauge group, $SU(2)_A \times SU(2)_B \times U(1)_Y$~\cite{Batra:2003nj, Lodone:2010kt}. We start considering both $H_{u}$ and $H_d$ to be charged under $SU(2)_A$; we will comment in the following on the chiral model~\cite{Craig:2013fga} where $H_u$ and $H_d$ are charged under the two different $SU(2)$.

The breaking $SU(2)_A \times SU(2)_B \rightarrow SU(2)_L$  is driven by a bidoublet $\Sigma$, which we parametrize as
\begin{equation}
\Sigma = \frac{1}{\sqrt{2}} \left( \sigma {\bf 1}_2 +  T^A \sigma^A\right) \, .
\end{equation}
The $\sigma$ and $T^A$ fields are a (complex) $SU(2)_L$ singlet and triplet, normalized to have canonical kinetic terms.

We add a singlet $S$ to the particle content to guarantee the breaking of the extended gauge symmetry also in the limit of exact SUSY. The most general superpotential is thus
\begin{equation}
 W = \mu H_u H_d + \lambda S \left( {\rm det} \Sigma -w^2 \right) +\lambda_S H_u H_d S + M_S S^2 + k S^3  \, .
\end{equation}
To simplify our discussion on the Higgs mass and couplings we assume $ \lambda_S  \lesssim 0.5 $. We checked that this choice gives negligible contributions to Higgs physics also for low $ \tan{ \beta } $.
For simplicity we also neglect the contributions from $M_S $ and $k$, considering the same superpotential as in \cite{Batra:2003nj}.
The modified $D$-terms are given by
\begin{equation}\label{eq:non_abelian_Dterms}
 \begin{array}{ccl}
  \displaystyle D^a_A &=& \displaystyle g_A \left( H_u^\dag \tau^a H_u + H_d^\dag \tau^a H_d + {\rm tr}(\Sigma^\dag \tau^a \Sigma) \right)\, ,\\
  \displaystyle D^a_B &=& \displaystyle g_B {\rm tr}( \Sigma \tau^a \Sigma^\dag) \, ,\\
  \displaystyle D_Y &=& \displaystyle g' \left( \frac{1}{2} |H_u|^2 - \frac{1}{2} |H_d|^2\right) \, ,\\
 \end{array}
\end{equation}
with the new gauge couplings satisfying $\frac{1}{g_A^2} + \frac{1}{g_B^2} = \frac{1}{g^2}$. The soft SUSY breaking potential is
\begin{equation}
 V_{SSB} = m^2_{H_u} |H_u|^2 + m^2_{H_d} |H_d|^2 + m^2_\Sigma |\Sigma|^2 + m^2_S |S|^2 + B H_u H_d  -B_\Sigma {\rm det}\Sigma + h.c.
\end{equation}
The $SU(2)_A \times SU(2)_B \times U(1)_Y \rightarrow SU(2)_L \times U(1)_Y$ breaking is driven by the singlet vev $\langle \sigma \rangle = u$, 
while EWSB is driven by $\langle H_u^0 \rangle = v_u$, $\langle H_d^0 \rangle = v_d$ and $\langle T^3 \rangle = v_T$. Notice that the triplet 
vev $v_T$ is bounded by EW precision measurements to satisfy $v_T \lesssim 3$ GeV, and is therefore negligible; we will comment on EWPM bounds in Sec.~\ref{sec:extra_gauge_bosons}.

Let us now compute the tuning associated with the $u$ and $v$ scales. In the $u \gg v \gg v_T$ limit, the minimum equations are
\begin{equation}\label{eq:non_abelian_minimum}
\begin{array}{ccl}
  {\displaystyle v_S} &=& {\displaystyle 0} \\
  {\displaystyle u^2} &=&  {\displaystyle\frac{2(B'_\Sigma - m^2_\Sigma)}{\lambda^2} \, , ~~~~~B'_\Sigma =B_\Sigma + \lambda^2 w^2 \, ,}\\
  {\displaystyle v_T} &=& {\displaystyle- \frac{g_A^2 \,  u\,  v^2 c_{2\beta}}{2  \left(4 m^2_\Sigma+ ( g_A^2 + g_B^2 ) u^2 \right)} \, ,}\\
  {\displaystyle v^2} &=& {\displaystyle - \frac{4\left( (m^2_{H_d} +  \mu^2) c_{\beta}^2 -(m^2_{H_u}+ \mu^2) s_\beta^2\right) }{(g^2 \eta  +g'^2) c^4_\beta -(g^2 \eta  +g'^2 +8 \lambda_u) s^2_\beta }  \, ,} \\
  {\displaystyle \frac{2B}{s_{2\beta}}} &=& {\displaystyle \frac{\left(g^2 \eta + g'^2 \right) \left(m^2_{H_u} + m^2_{H_d} + 2 \mu^2 \right) c_{2\beta} - 8 \lambda_u s^2_\beta \left(m^2_{H_d} +  \mu^2 \right)}{(g^2 \eta  +g'^2) c^4_\beta -(g^2 \eta  +g'^2 +8 \lambda_u) s^2_\beta }\, ,}
 \end{array}
\end{equation}
where
\begin{equation}\label{eq:eta}
 {\displaystyle \eta = \frac{1+ \frac{4 (m^2_\Sigma/u^2)}{g_B^2}}{1+ \frac{4 (m^2_\Sigma/u^2)}{g_A^2+g_B^2}}\, .}
\end{equation}
To account for loop corrections,  we have introduced a quartic term $\lambda_u |H_u^0|^4$ in the scalar potential. We will compute in detail $\lambda_u$ in the following, but for the moment we will remain agnostic about its form.

The computation of the tuning on $u^2$ gives
\begin{equation}\label{eq:tuning_u^2}
 \Delta^{u^2}_{w^2} = \frac{4 w^2 }{\lambda^2 u^2}\, ,~~~~~\Delta^{u^2}_{m^2_\Sigma} = \frac{2 m^2_\Sigma }{\lambda^2 u^2}\,,  ~~~~~ \Delta^{u^2}_{B_\Sigma} = \frac{2 B_\Sigma }{\lambda^2 u^2}\, .
\end{equation}
Requiring $u^2$ to be basically untuned ({\it i.e.} $\Delta^{u^2} < 5$), we get 
\begin{equation}\label{eq:bound_FT_u2}
 \frac{w^2}{u^2} \lesssim \frac{5\lambda^2}{4}\, , ~~~~~\frac{m^2_\Sigma}{u^2} \lesssim \frac{5\lambda^2}{2}\, , ~~~~~\frac{B_\Sigma}{u^2} \lesssim \frac{5\lambda^2}{2}\, .
\end{equation}
Notice that the second inequality is particularly important, since it sets a natural bound on the ratio $m_\Sigma^2/u^2$ that appears in $\eta$, Eq.~(\ref{eq:eta}), and that cannot be inferred from the low energy theory. All the upper bounds depend on the $\lambda$ coupling. Solving the relevant RGE's~\cite{Lodone:2010kt}, we find that for $\lambda \lesssim 1.2$ at low energy, the coupling remains perturbative up to the Planck scale.\\

Turning to the tuning on the EW scale, the relevant sensitivities are given by
\begin{equation}\label{eq:tuning_EW_DMSSM}
  \begin{array}{ccl}
    \displaystyle \Delta^{v^2}_{m^2_{H_d}} &=& \displaystyle \left| \frac{m^2_{H_d}}{v^2} \frac{\delta v^2}{\delta m^2_{H_d}} \right|\, , \\[0.3cm]
    \displaystyle \Delta^{v^2}_{\mu} &=& \displaystyle \left| \frac{\mu}{v^2} \frac{\delta v^2}{\delta m^2_{H_u}}\left[ 2\mu - 2 v^2 \frac{\delta \lambda_u}{\delta\mu} \right]\right| \, , \\[0.3cm]
    \displaystyle \left. \Delta^{v^2}_{m^2_\Sigma} \right|_{\rm tree} &=& \displaystyle \left|  \frac{4 g_A^4 (m_{\Sigma }^2/u^2)}{ \left(g_A^2+g_B^2\right) \left(g_A^2 \left(g_B^2+g'^2+8 \lambda _u\right)+g_B^2   \left(g'^2+8 \lambda _u\right)\right)}\right| \, , \\[0.3cm]
    \displaystyle \left. \Delta^{v^2}_{m^2_\Sigma} \right|_{\rm loop} &=& \displaystyle \left| \frac{m^2_\Sigma}{v^2} \frac{\delta v^2}{\delta m^2_{H_u}} \frac{\delta m^2_{H_u}}{\delta m^2_\Sigma} \right| \, , \\[0.3cm]
    \displaystyle \Delta^{v^2}_{\xi_{\tilde t}}&=& \displaystyle \left| \frac{\xi_{\tilde t}}{v^2} \frac{\delta v^2}{\delta m^2_{H_u}} \left[ \frac{\delta m^2_{H_u}}{\delta \xi_{\tilde{t}}} -  2  v^2 \frac{\lambda_u}{\delta \xi_{\tilde{t}}}\right]\right|  \, ,
  \end{array}
\end{equation}
where $\delta v^2 / \delta m^2_{H_u}$ and $\delta v^2 / \delta m^2_{H_d}$ can be inferred from Eq.~(\ref{eq:FT_general}), while $\xi_{\tilde t}= \big\{ m^2_{\tilde{Q}_3},m^2_{\tilde{t}_3}, A_t \big\} $ are the stop parameters.

The soft SUSY breaking mass $m^2_\Sigma$ appears in the minimum equations at tree level through $\eta$, and at the two loop level in $m^2_{H_u}$. The tree level bound only constraints the ratio $m^2_\Sigma/u^2$, while the relevant RGE to be taken into account for the computation of $\left. \Delta^{v^2}_{m^2_\Sigma} \right|_{\rm loop}$ is~\cite{Lodone:2010kt}
\begin{equation}
\frac{d m^2_{H_u}}{d\log Q} = \frac{6}{(16\pi^2)^2} g_A^4 m^2_\Sigma \, .
\end{equation}
Since $g_A$ changes significantly with the scale, we properly integrate its RGE in our estimate of the fine tuning.
\begin{figure}[tb]
 \begin{center}
  \includegraphics[width=.46\textwidth]{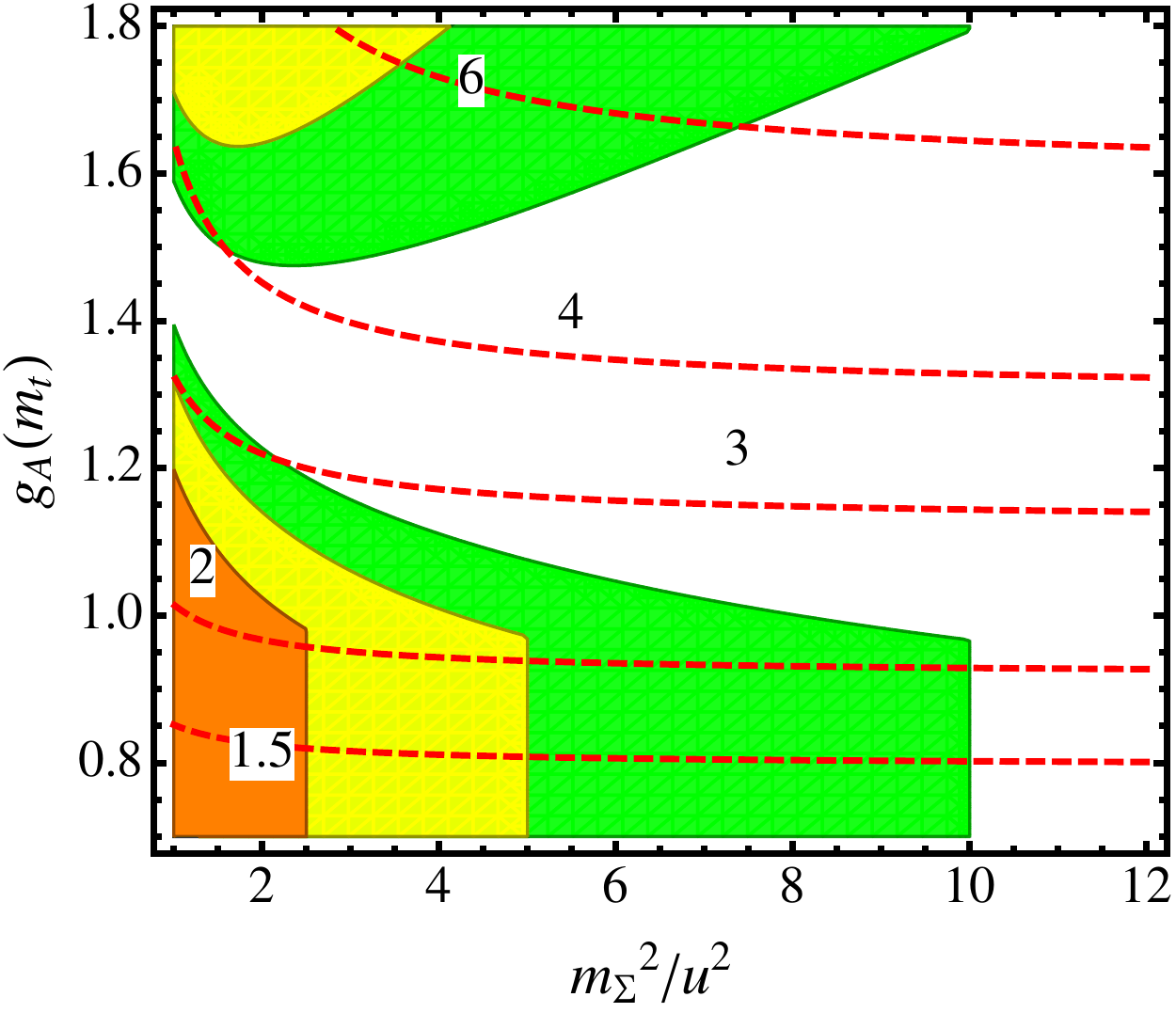} 
  \includegraphics[width=.48\textwidth]{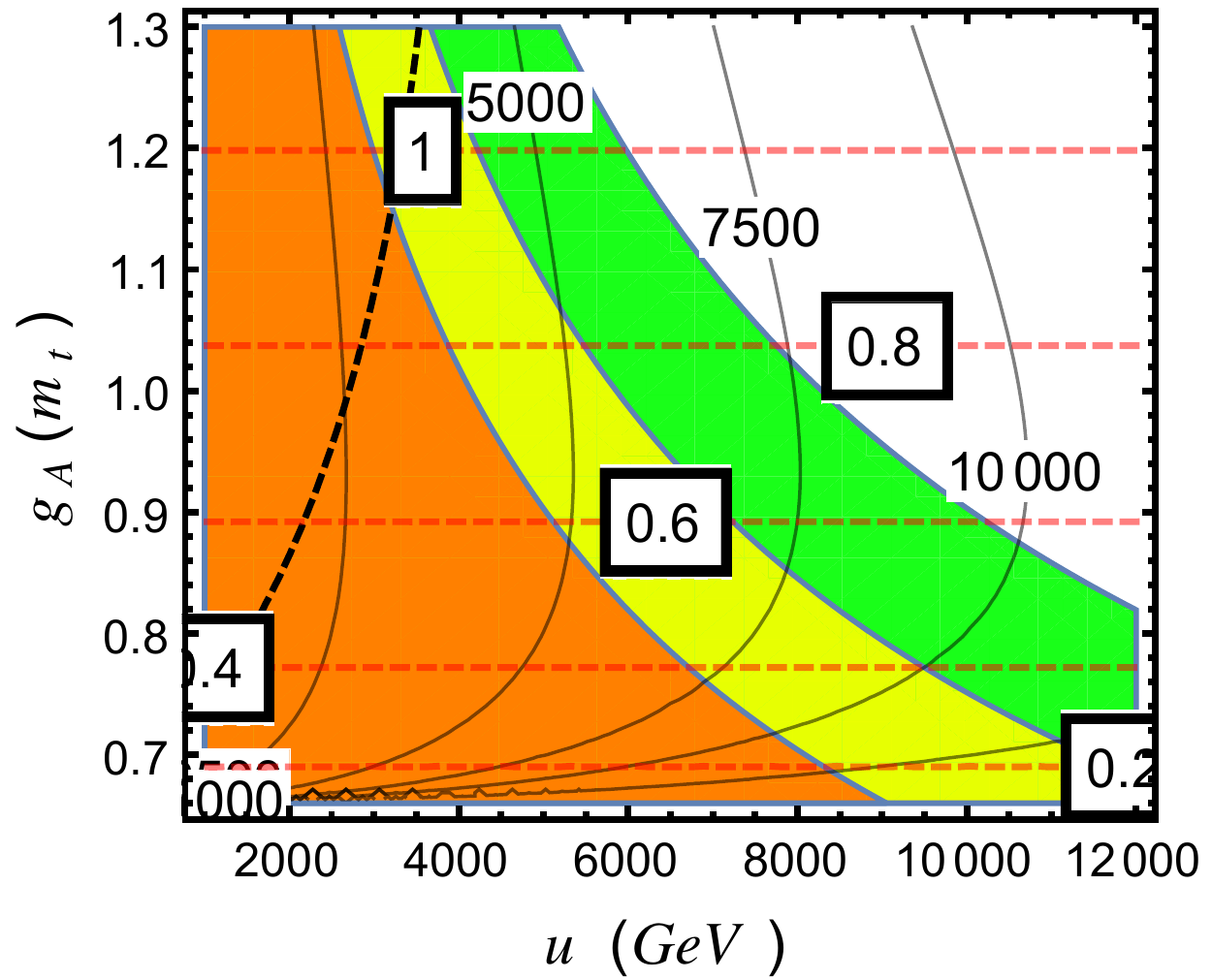}
 \end{center}
 \caption{\label{fig:gauge_masses} Left panel: contours of $\eta$, Eq.~(\ref{eq:eta}), as a function of $m^2_\Sigma/u^2$ and $g_A(m_t)$. The orange, yellow and green regions refer to ${\rm max}\left( \Delta^{u^2}_{m^2_\Sigma}, \left.\Delta^{v^2}_{m^2_\Sigma}\right|_{tree} \right) <5$, $10$ and $20$ respectively. Right panel: contours of $m_{W'}$, Eq.~(\ref{eq:gauge_boson_mass}), as a function of $u$ and $g_A(m_t)$. Colored regions as in left panel, for $\Delta_{u^2}$. Red lines (framed labels): contours of $g_{W'}$, Eq.~(\ref{eq:universal_coupling}). Dashed black line: current bound from EWPM.}
\end{figure}
In Fig.\ref{fig:gauge_masses} (left panel) we show contours of $\eta$ in the $ (m_\Sigma^2/u^2, g_A)$ plane, together with the  tuning on $m^2_\Sigma/u^2$, Eqs.~(\ref{eq:bound_FT_u2})-(\ref{eq:tuning_EW_DMSSM}). The orange, yellow and green regions refer to $\Delta<5\, , 10$, and $20$, respectively. We see that for moderate values of the gauge coupling $g_A$, values $\eta\lesssim3$ are compatible with a tuning better than $20\%$. We will show in Sec.~\ref{sec:DMSSM_Higgs_mass} that such values can accommodate a $126\,{\rm GeV}$ Higgs without relying on large radiative corrections from the stop sector.
Furthermore, naturalness requires $ g_A(m_t) \lesssim 1.2$, which is compatible with the request of perturbativity up to the GUT scale. In the present paper we take a bottom-up approach, and  we refer to~\cite{Batra:2003nj} for comments about unification in these models. We plan to address this issue in a future work.

The expression for $\left. \Delta^{v^2}_{m^2_\Sigma} \right|_{\rm loop}$ can be used to place a naturalness bound on the absolute SUSY breaking scale of the bidoublet. Insisting on $\Delta_{m^2_\Sigma} <5$, we find $m_\Sigma \lesssim 6.5$ TeV. Assuming $ \lambda \sim 1$, this translates into a naturalness bound for $ u$, $ u \lesssim 4.1$ TeV.

\subsection{Extra gauge bosons as new signals of naturalness}\label{sec:extra_gauge_bosons}

As we have argued in the previous section, it is possible to have a natural EW scale as long as the additional scalars that break the extended gauge symmetry have mass $m_\Sigma \lesssim  6.5$ TeV. At the same time, the $u$ scale is itself natural for $m_\Sigma/u \lesssim \sqrt{5/2}$, Fig.~\ref{fig:gauge_masses}, in such a way that for $u \lesssim 4$ TeV the tuning due to the extended gauge sector is never worse than $20\%$. Since $u$ sets the mass scale of the heavy gauge bosons, we conclude that these states are likely to be observed at the LHC-13. Let us make the argument more concrete.

In the $u \gg v \gg v_T$ limit, the gauge boson masses simplify to
\begin{equation}\label{eq:gauge_boson_mass}
 \begin{array}{ccl}
  \displaystyle m^2_W &\simeq & \displaystyle \frac{g^2 v^2}{2} \left[ 1 - \left(\frac{g_A}{g_B}\right)^4 \frac{v^2}{u^2} + 4 \frac{v_T^2}{v^2} \right]  \, ,\\
  \displaystyle m^2_Z &\simeq & \displaystyle \frac{(g^2+g'^2) v^2}{2} \left[ 1 - \left(\frac{g_A}{g_B}\right)^4 \frac{v^2}{u^2} \right] \, ,\\
  \displaystyle m^2_{W'} &\simeq & \displaystyle \frac{(g_A^2 + g_B^2) u^2}{2} \left[ 1+\left(\frac{g_A}{g_B}\right)^4 \frac{v^2}{u^2} + \left( \frac{g_A^2 - g_B^2}{g_A^2+g_B^2} \right)^2 \frac{v_T^2}{u^2}\right] \, ,\\
  \displaystyle m^2_{Z'} &\simeq & \displaystyle \frac{(g_A^2 + g_B^2)  u^2}{2} \left[ 1+\left(\frac{g_A}{g_B}\right)^4 \frac{v^2}{u^2} +\frac{v_T^2}{u^2} \right] \, ,
 \end{array}
\end{equation}
with the two heavy vectors basically degenerate, and $\rho$ parameter given by $\rho \simeq 1 + 4 \frac{v_T^2}{v^2}$.

The heavy gauge bosons couple to the SM doublets charged under $SU(2)_A$ with universal strength (see Appendix~\ref{sec:appendix_EWPM})
\begin{equation}\label{eq:universal_coupling}
g_{W'} = g \frac{g_A}{g_B}\, .
\end{equation}
An analogous expression can be derived for the $SU(2)_B$ doublets, with the replacement $g_A/g_B \rightarrow g_B/g_A$. We consider here a scenario in which all SM doublets are charged under $SU(2)_A,$ therefore coupling to the heavy gauge bosons with strength given by Eq.~(\ref{eq:universal_coupling}).
Another interesting possibility~\cite{Batra:2003nj} is to charge the first and second generation under $SU(2)_B$ and the third one under $SU(2)_A$. This makes $SU(2)_A$ asymptotically free, with larger values of $g_A$ (and of $\eta$ as well, see Eq.~(\ref{eq:eta})) allowed at the EW scale. However, naturalness does not allow for arbitrarily large values of $g_A$ and $\eta$, as can be seen from the left panel of Fig.~\ref{fig:gauge_masses}. Since, as already anticipated, $\eta$ is related to the enhanced tree level Higgs boson mass, it is clear that we cannot obtain an arbitrarily heavy Higgs without worsening the fine tuning. In any case, modest values of $\eta$ are sufficient to accommodate a $126$ GeV Higgs (see Sec.~\ref{sec:DMSSM_Higgs_mass}), in a such a way that the Higgs boson mass does not require $SU(2)_A$ to be asymptotically free to agree with experiments with a low fine tuning. 
Still, it may be worth to explore such a scenario because it could offer a valuable starting point to build a UV completion for spectra with the first and second generation squarks heavier than the third one~\cite{Batra:2003nj, Dimopoulos:2014aua}.\\

Let us now turn to the interplay between naturalness and the heavy gauge boson masses, as shown in Fig.~\ref{fig:gauge_masses} (right panel). The masses of the triplet of heavy gauge bosons are shown as black continuous lines (unframed labels), while the orange, yellow and green regions refer to $\Delta_{u}<5$, $10$ and $20$, respectively. We also show contours of the universal coupling $g_{W'}$ defined in Eq.~(\ref{eq:universal_coupling}) (red dashed lines, framed labels), as well as the bound coming from Electroweak Precision Measurements (EWPM) (black dashed line). We discuss in detail how we obtain this bound in Appendix~\ref{sec:appendix_EWPM}. 

We see that for $g_A\simeq g$, even a $15$ TeV gauge boson does not introduce any relevant tuning in the DMSSM. Moreover, such a heavy state will likely escape detection at the LHC, since $g_{W'} \simeq 0.2$ in this region. Notice however that this portion of parameter space is disfavored by the Higgs boson mass: since $\eta \simeq 1$ (Fig.~\ref{fig:gauge_masses}, left panel), we are effectively in the MSSM limit of the DMSSM, with the usual fine tuning problems related to the Higgs boson mass.

On the contrary, for $g_A\simeq g_B \simeq (0.9 \div 1)$, $\eta$ is large enough to ensure that the Higgs mass can be accommodated without introducing any relevant tuning. For these values, the requirement of a tuning better than $20\%$ leads to an upper bound $m_{W'} \lesssim 6$ TeV, and since $g_{W'} \simeq g$, it is not unlikely that these states can be detected at the LHC-13~\cite{Godfrey:2013eta}. A 100 TeV collider would certainly be an ideal ground to explore naturalness in this framework. \\

Turning to the direct searches at hadron colliders, it is worth  to point out that, unlike what happens in non-SUSY extended gauge sectors, here decays into light superpartners such as the squarks of the third generation can be relevant, and must be taken into account in the study of the phenomenology. We postpone a detailed analysis of the issue to a future work.

\section{ A natural 126 GeV SM-like Higgs from non-decoupling $D$-terms}

In this section we show that, using the non decoupling $D$-terms, the mass of the lightest CP-even scalar can be raised up to $126$ GeV without requiring any relevant tuning in a large region of parameter space. The same can be achieved in other extension of the MSSM as well, such as the NMSSM~\cite{Gherghetta:2012gb}. However, since in this case small values of $\tan\beta$ are required, generically the Higgs results SM-like only for quite heavy scalars. As we saw in Sec.~\ref{sec:tools}, this may introduce a relevant source of tuning. Higgs couplings measurements at the Run II of the LHC can already probe a fine tuning at a few percent level~\cite{Gherghetta:2014xea}. On the contrary, we will show that in the DMSSM the lightest CP-even Higgs is SM-like in a large region of the parameter space, due to the possibility of considering moderate or large values of $\tan{\beta}.$ In other words, in this scenario the decoupling limit $ m_A \gg m_h$ is natural.

\subsection{126 GeV Higgs in the DMSSM}\label{sec:DMSSM_Higgs_mass}

Let us now explain in detail how in the DMSSM the tree level Higgs quartic coupling is increased with respect to the MSSM. From Eq.~(\ref{eq:non_abelian_Dterms}) we see that once the EW singlet $\sigma$ acquires its vev, the bidoublet gets shifted, $\Sigma\rightarrow \frac{u}{\sqrt{2}}{\bf 1} + \Sigma$, generating a trilinear coupling~\cite{Huo:2012tw}
\begin{equation}
V_D \supset \frac{g_A^2 u}{2\sqrt{2}} \left( H_u^\dag \tau^A H_u +  H_d^\dag \tau^A H_d \right) \frac{T^A + \bar{T}^A}{\sqrt{2}}  + \dots
\end{equation}
Since the real scalar triplet $T^A_R = \frac{T^A + \bar{T}^A}{\sqrt{2}}$ is always heavy, $m^2_{T^A_R} = 2 m^2_\Sigma + \frac{1}{2} \left(g_A^2 + g_B^2\right) u^2$, it can be integrated out, generating the effective $D$-terms 
\begin{equation}
\label{ Vde}
V_D^{eff} = \frac{g^2 \eta}{2} \left( H_u^\dag \tau^A H_u + H_d^\dag \tau^A H_d\right)^2 + \frac{g'^2}{2}\left( \frac{1}{2}|H_u|^2 - \frac{1}{2} |H_d|^2\right) \, . 
\end{equation}
This is a general result, and can be applied with straightforward modifications to any gauge extended SUSY model. Indeed, whenever the fields driving symmetry breaking acquire a vev, a trilinear coupling is always generated, leaving at low energy effective $D$-terms that can be parametrized as
\begin{equation}\label{eq:general_mod_Dterm}
\begin{array}{ccl}
V_D^{eff} &=& \frac{g^2 \eta}{2} \left( H_u^\dag \tau^A H_u + H_d^\dag \tau^A H_d \right)^2  \\
&& \mbox{} + \frac{g'^2 \eta'}{2} \left( \frac{1}{2}|H_u|^2 - \frac{1}{2} |H_d|^2 \right)^2 \, .
\end{array}
\end{equation}
where $ \eta $ and $\eta'$ have different expressions depending on the concrete realization under consideration. 

The tree level mass matrix for the CP-even scalars is easily computed. Rotating to the vev basis $(h, H)$, we have
\begin{equation}\label{eq:general_M2}
 {\cal M}^2 = 
 \begin{pmatrix}
  (g^2 \eta + g'^2 \eta') \frac{v^2}{2} c_{2\beta}^2 & (g^2 \eta + g'^2 \eta') \frac{v^2}{4} s_{4\beta} \\ 
  (g^2 \eta + g'^2 \eta') \frac{v^2}{4} s_{4\beta} \ & (g^2 \eta + g'^2 \eta') \frac{v^2}{2} s_{2\beta}^2 + \frac{2B}{s_{2\beta}} 
 \end{pmatrix}\, .
\end{equation}
We clearly see that for large enough $\eta$ and $\eta'$ the mass of the lightest scalar is larger than the $Z$ boson already at tree level.\\

In order to precisely estimate the value of $\eta$ needed to accommodate $m_h \simeq 126$ GeV, we turn now to the computation of the Coleman Weinberg potential. To this purpose it is sufficient to consider the effective theory below the real heavy scalars threshold, which we always assume $ \gtrsim 1$ TeV due to constraints from EWPM.\\

Let us start considering the scalar contributions. For simplicity, we will neglect the down-type Yukawa couplings for all the three generations. In addition, we compute the eigenvalues of the mass matrices appearing in the Coleman Weinberg potential expanding in powers of $H_u^0$. Taking the LH and RH masses for the squarks of the first two generations to be degenerate together with the RH sbottom, the Higgs quartic coupling at one loop is given by
\begin{equation}\label{eq:general_CW}
 \begin{array}{ccl}
  V_{CW}^s & \supset &   \bigg[\frac{9}{96\pi^2}  h_t^4 \log\frac{m^2_{t_R}}{Q^2} 
			+ \frac{3}{128\pi^2}  \delta_{L12} (g^2 \eta)^2  \log\frac{m^2_{12}}{Q^2} \\
	  && \phantom{\bigg[} \mbox{} +\frac{\left( 12 h_t^2 -3 \delta_{L3} g^2 \eta \right)^2 + 
		       \left( 3 \delta_{L3} g^2 \eta \right)^2}{1536\pi^2} \log\frac{m^2_{Q_L}}{Q^2} \\
	  && \phantom{\bigg[} \mbox{} - \frac{h_t^2 A_t^2}{64\pi^2} \left( \frac{3 g^2 \eta^2 - 5 g'^2}{m_{\tilde{Q}_L}^2 - m_{\tilde{t}_R}^2}- \frac{12 h_t^2 A_t^2}{(m_{\tilde{Q}_L}^2 - m_{\tilde{t}_R}^2)^2}\right)  \bigg] |H_u^0|^4 \, .   
 \end{array}
\end{equation}
Here $\delta_{Li}$ refers to the $i^{th}$ generation, and $\delta_{Li}=1$ only if the squarks of the corresponding generation are doublets under $SU(2)_A$ as the two Higgses; otherwise $\delta_{Li}=0$. 
We checked that the $H_d$ contribution, which in principle should also be included, is always subleading with respect to the squark one.\\

Turning to the fermionic contribution, we consider only the neutral and charged Higgsinos in the effective theory. This is a good approximation, since the calculation of the neutralinos and charginos mass matrix shows that all the other fermions receive an irreducible ${\cal O}(u)$ contribution to their masses. There may be important mixing effects proportional to the gaugino mass pushing down some eigenvalue (see \cite{Huo:2012tw} for interesting phenomenological consequences in the Higgs sector);  we checked numerically that our approximation is reliable in a broad region of parameter space. The Higgsino contribution to the Coleman Weinberg potential is given by
\begin{equation}
V_{CW}^f \supset -\left[ \frac{(g_A^2+g'^2)^2}{128 \pi^2} \log\frac{\mu^2}{m_t^2} + \frac{g_A^4}{16\pi^2} \log\frac{\mu^2}{m_t^2} \right] |H_u^0|^4\, ,
\end{equation}
where the first and second term come from the neutral and charged Higgsinos, respectively. Since naturalness requires $\mu$ to be fairly light, we do not expect this contribution to give a significant reduction to the Higgs quartic coupling; nevertheless, we have included it in our numerical study.\\

In order to make more reliable our computation of the lightest CP-even mass, we evolve the Higgs quartic coupling from its boundary, $\lambda(\Lambda)= \frac{g^2(\Lambda) \eta + g'^2(\Lambda)}{4} \cos^2{2 \beta}$,~\footnote{We use the convention $V\supset \frac{\lambda}{2} |H|^4$, with the Higgs boson mass given by $m_h^2 = 2 \lambda v^2$.} down to $m_t$, taking into account the different thresholds encountered evolving from high to low energy. We will consider a simplified situation in which the stops are generate at $m_{\tilde t}$, with the hierarchy $m_{12} \gg m_{\tilde t} \gg \mu$. We can thus write:
\begin{equation}\label{eq:quartic_total}
 \lambda(m_t)  \simeq  \frac{g^2(\Lambda) \eta + g'^2(\Lambda)}{4}  \cos^2{2 \beta}+\bigg( \lambda_{\tilde{q}_{12}}(m_{12}) 
 + \delta\lambda_{\tilde t}(m_{\tilde t})- \delta\lambda_{\chi^0}(\mu) \bigg) \sin^4{ \beta}\\
\end{equation}
where, assuming $\delta_{L3}=1$ and $\delta_{L1,2}=1$ in Eq.~(\ref{eq:general_CW}), we have
\begin{equation}
 \begin{array}{ccl}
  {\displaystyle \delta \lambda_{\tilde{q}_{12}}} &=& {\displaystyle \frac{3g^4 \eta^2}{64\pi^2}  \log\frac{m^2_{12}}{m_t^2}  \, ,}\\
  {\displaystyle \delta\lambda_{\tilde t}} &=& {\displaystyle\left(\frac{9 h_t^4}{48\pi^2}  + \frac{(12 h_t^2 -3 g^2 \eta)^2 + 9 g^4 \eta^2}{768\pi^2} \right) \log\frac{m^2_{\tilde t}}{m_t^2} +\frac{h_t^2 A_t^2}{128\pi^2} \left(\frac{3 (8 h_t^2 - g^2 \eta - g'^2)}{m^2_{\tilde t}} -  \frac{2 h_t^2 A_t^2}{m^4_{\tilde t}}\right)} \\
		      & & {\displaystyle \mbox{} + \frac{6 h_t^2}{(16\pi^2)^2} \bigg( \frac{3}{2} h_t^2 - 32 \pi \alpha_3(m_t) \bigg) \log^2\frac{m^2_{\tilde t}}{m_t^2} \, , }\\
{\displaystyle \delta\lambda_{\chi^0} } &=& {\displaystyle \frac{9 g_A^4 + 2 g_A^2 g'^2 +g'^4}{128 \pi^2} \log\frac{\mu^2}{m_t^2} \, .}
 \end{array}
\end{equation}
All the couplings are evaluated at the relevant scale, {\it i.e.} $m_{12}$ for $\delta \lambda_{\tilde{q}_{12}}$, $m_{\tilde t}$ for $\delta \lambda_{\tilde{t}}$ and $\mu$ for $\delta\lambda_{\chi^0}$. Notice that we also take into account the two loops contributions from the stop system, since they can give a sizable negative contribution.\\
\begin{figure}[tb]
 \begin{center}
  \includegraphics[width=.55\textwidth]{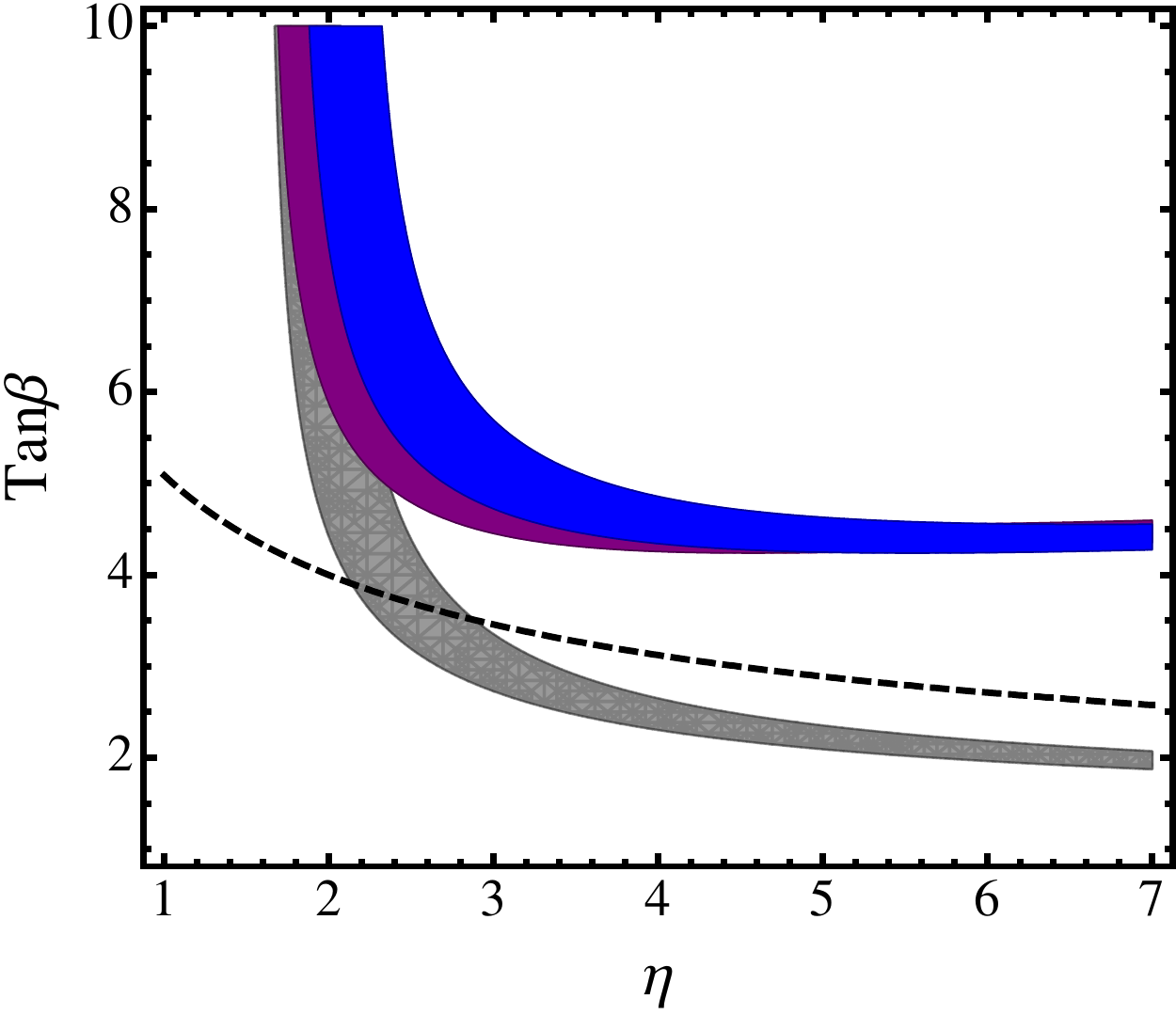} 
 \end{center}
 \caption{\label{fig:Higgs_massT} Regions in which $115$ GeV $<(m_h)^{tree}  <126 $ GeV, Eq.~(\ref{eq:general_M2}), as a function of  $ \tan{\beta} $ and $ \eta$, for $m_A=200$ GeV (blue and purple regions) and $m_A=700$ GeV (gray region).
 We take $ \eta=\eta'$ for the blue and gray regions, and $\eta'=1$ for the purple one. The black dashed  line corresponds to the contour for $ \Delta_{m_A} =5$ for $ m_A=700$ GeV.}
\end{figure}

We can now discuss the region of parameter space in which a $126$ GeV Higgs is obtained. To this end, we start requiring the maximum tuning coming from the stop sector to be no worse than $20\%$, {\it i.e.} ${\rm max}( \Delta_{X_t}, \Delta_{m^2_{\tilde t}}) < 5 $. This sets an upper bound on the size of the stops radiative contribution, that in turn translates into a lower bound for the tree level Higgs mass. We find this to be $m_h^{tree} \gtrsim 115$ GeV. Notice that we do not consider the running of the stop mass parameters in the computation of the tuning, since it is highly dependent on the value of the gluino mass. Whenever we will talk about tuning on the stop sector, it has to be interpreted as the worst possible tuning, arising for vanishing gluino mass.\\

We show in Fig.~\ref{fig:Higgs_massT} the $115$ GeV $< m_h^{tree} < 126$ GeV region in the $(\eta, \tan\beta)$ plane, for $m_A=200$ GeV (blue and purple regions) and $m_A=700$ GeV (gray region). We also show the contours of $\Delta_{m_A^2} = 5$ for the two values of $m_A$ considered (dashed line and continuous line for $m_A=700$ GeV and $200$ GeV, respectively). In the diagonalization of Eq.~(\ref{eq:general_M2}), we have fixed $\eta'=1$ for the purple region, while $\eta'=\eta$ for the gray and blue regions. By construction, the tuning due to the stop system is not a problem in accommodating the Higgs boson mass. Moreover, we see that the tuning on $m_A$ sets a lower bound $\tan\beta \gtrsim 3$ for $m_A=700$ GeV, and basically no bound for $m_A=200$ GeV. Comparing with Fig.~\ref{fig:gauge_masses}, we also see that as long as $\eta\lesssim 3$, also the tuning associated with the ratio $m^2_\Sigma/u^2$ is under control. This allows us to conclude that as long as $\tan\beta \gtrsim 4$, the Higgs boson mass in the DMSSM can be easily accommodated with a tuning better than $20\%$. \\

Let us conclude with a comment on the tuning needed to stabilize the Higgs mass itself - {\it i.e.} the Higgs quartic coupling for fixed vev $v$. In models with an extended Higgs sector, this may represent a non trivial source of sensitivity, as is the case in the NMSSM for $\lambda \gtrsim 1$~\cite{Gherghetta:2012gb}. In our case, the correct Higgs boson mass is obtained essentially at tree level (with no need for relevant loop corrections) as long as the states belonging to the bidoublet, $\sigma$ and $T^3$, are sufficiently decoupled from the doublet system, in such a way that Eq.~(\ref{eq:general_M2}) applies. This can be achieved for large $m^2_\Sigma$, and as we have already seen, as long as $m_\Sigma \lesssim 6.5$ TeV the naturalness of the EW scale is not compromised. We can now ask whether the sensitivity of the Higgs boson mass on $m^2_\Sigma$ is instead increased, as we make it larger. This is not the case. Using the seesaw formula, we immediately see that since $m^2_\Sigma$ does not appear in the mixing terms, the tuning on the Higgs boson mass goes schematically with $\Delta \sim ({\rm mixing}/m^2_\Sigma)^2$.

\subsection{SM like Higgs couplings and fine tuning implications}\label{sec:Higgs_couplings}

Let us now investigate whether precision Higgs physics can provide a good test of naturalness in our framework, as in~\cite{Farina:2013ssa, Farina:2013fsa, Gherghetta:2014xea}. We study the modifications with respect to the SM couplings, $r_i = \sigma_i / \sigma_i^{SM}$, assuming that no new production or decay modes are present beside the usual SM ones.

At tree level, the mixing between $h$ and $H$ modifies the couplings to the SM vectors and fermions~\cite{Gherghetta:2014xea} as follows:
\begin{equation}
 \begin{array}{ccl}
  {\displaystyle r_u} & \simeq & {\displaystyle 1+\frac{1}{t_\beta} \left( \frac{{\cal M}^2_{hH}}{{\cal M}_{HH}^2} + \frac{{\cal M}_{hh}^2 {\cal M}_{hH}^2}{{\cal M}_{HH}^4}\right) -\frac{{\cal M}_{hH}^4}{2{\cal M}_{HH}^4} \, ,}\\
  {\displaystyle r_d} & \simeq & {\displaystyle 1-t_\beta \left( \frac{{\cal M}^2_{hH}}{{\cal M}_{HH}^2} + \frac{{\cal M}_{hh}^2 {\cal M}_{hH}^2}{{\cal M}_{HH}^4}\right) -\frac{{\cal M}_{hH}^4}{2{\cal M}_{HH}^4}\, ,} \\
  {\displaystyle r_V} & \simeq & {\displaystyle 1- \frac{{\cal M}_{hH}^4}{2{\cal M}_{HH}^4} \, ,}
 \end{array}
\end{equation}
where ${\cal M}^2_{hh}$, ${\cal M}^2_{hH}$ and ${\cal M}^2_{HH}$ are the matrix elements of Eq.~(\ref{eq:general_M2}).

At loop level, the main contributions to the coupling to gluons arise from stop and sbottom loops, while for the coupling to photons the lightest chargino may be relevant as well~\cite{Blum:2012ii}. In the case of a sbottom mixing of weak size, $\tilde{b}_{1,2}$ contribute with deviation at the percent level. Such deviations can however be made much smaller assuming small (or vanishing) mixing in the sbottom sector, or moderately heavy sbottoms. In the following we will always assume the sbottoms to be heavy enough to suppress this contribution (we explicitly checked that already for $m_{\tilde{b}} \gtrsim 500$ GeV the deviations in $r_G$ are below the $5$ permil level). On the contrary, even for large $\tan\beta$, chargino mediated deviations in the $h\gamma\gamma$ coupling cannot be made smaller than $\pm (1-2) \%$~\cite{Blum:2012ii}. Their interference with the stop contribution, while not particularly relevant with the current precision on the $r_\gamma$ measurement, will become important when a percent precision will be reached at future colliders.

We can parametrize the deviations from the SM due to stops as
\begin{equation}
 r_{GG} = |1+ \delta r_G|^2 \, ,
 ~~~~
 r_{\gamma\gamma} \simeq |1- 0.27 \delta r_G|^2 \, ,
\end{equation}
where
\begin{equation}
 \delta r^{\tilde t}_G \simeq \frac{m^2_t}{4} \left( \frac{1}{m_{\tilde{t}_1}^2} + \frac{1}{m_{\tilde{t}_2}^2} - \frac{X_t^2}{m_{\tilde{t}_1}^2 m_{\tilde{t}_2}^2} \right) \, ,
\end{equation}
with $m_{\tilde{t}_{1,2}}$ the physical stop masses.\\
\begin{figure}[tb]
 \begin{center}
  \includegraphics[width=.45\textwidth]{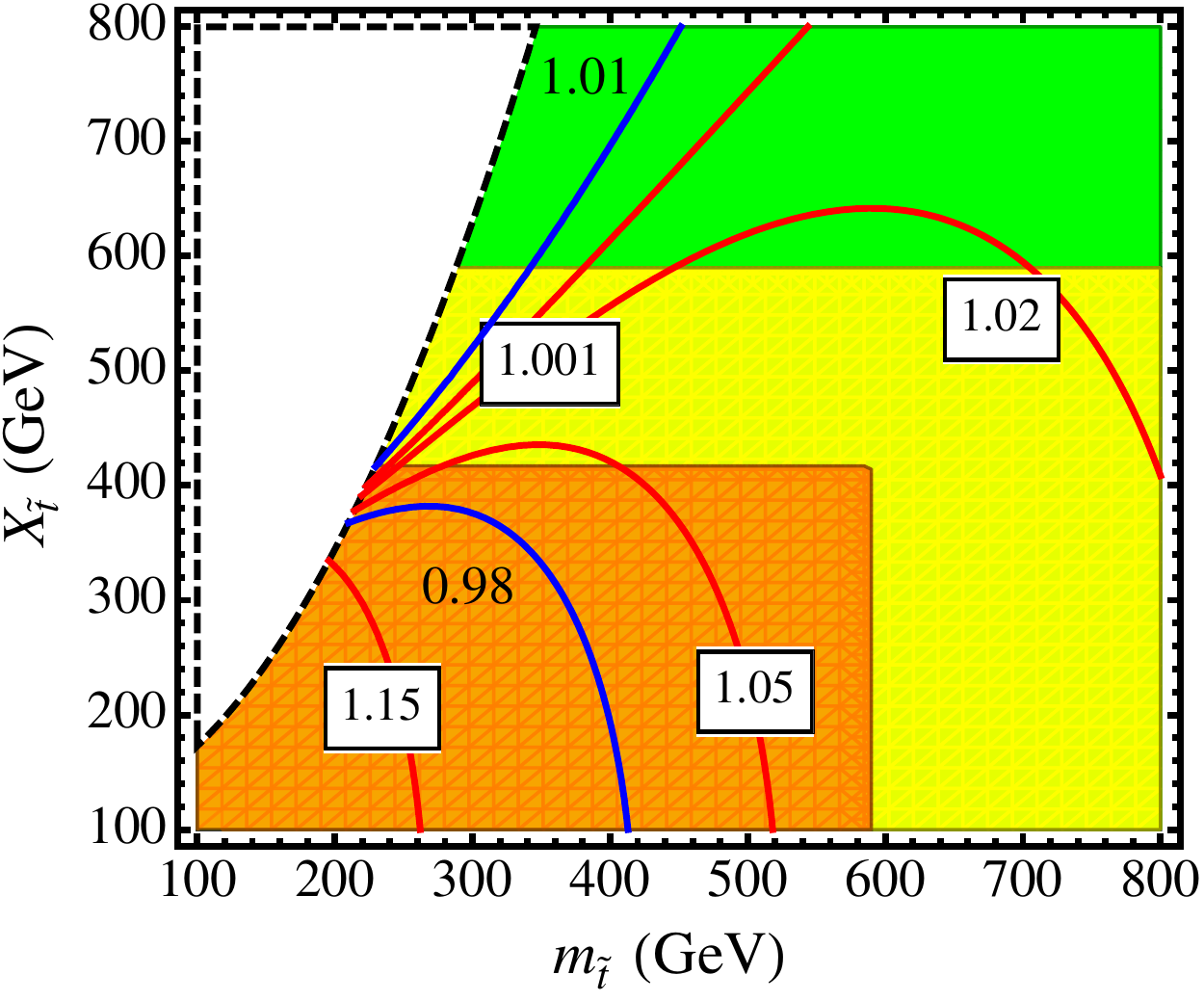} 
   \includegraphics[width=.48\textwidth]{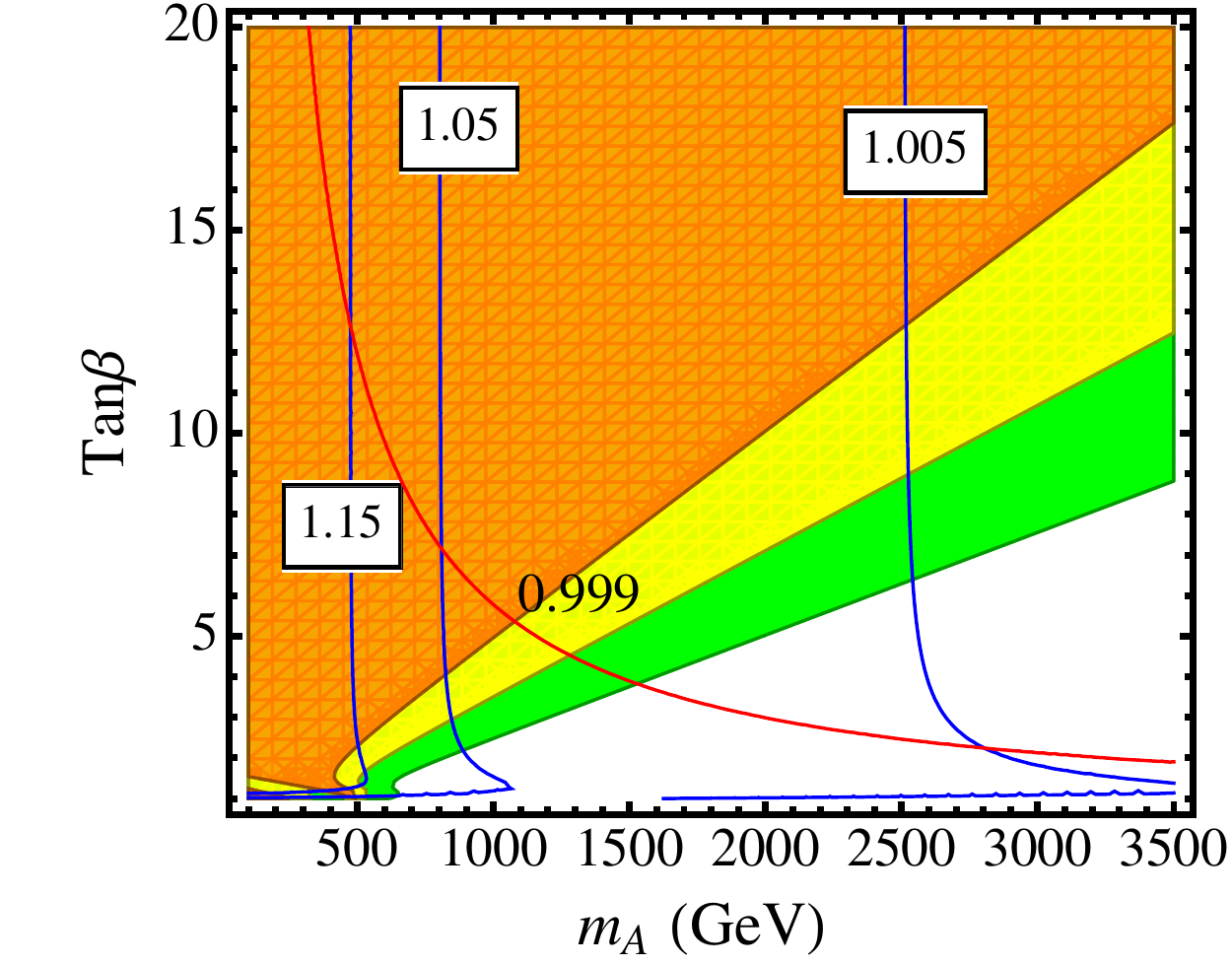} 
 \end{center}
 \caption{\label{fig:HiggsC} Left panel: contours of $r_{GG}$ (red lines) and $r_{\gamma\gamma}$ (blue lines). The white region is excluded requiring positive stop-loop contribution to the Higgs quartic and charge/color conservation. The orange, yellow and green regions refer to $\Delta_{\tilde t}<5$, $10$ and $20$, respectively. Right panel: contours of $r_u$ (red lines) and $r_d$ (blue lines). Same color code as the left panel for the fine tuning regions, this time referred to $\Delta_{m_A^2}$.}
\end{figure}
Let us now study what naturalness predicts for Higgs couplings deviations from the SM behavior. 
In Fig.~\ref{fig:HiggsC}, left panel, we show the contours of $r_{GG}$ (red lines) and $r_{\gamma\gamma}$ (blue lines) in the $(m_{\tilde t}, X_t)$ plane, assuming $m_{\tilde t} = m_{Q_{L3}} = m_{t_R}$, while in the right panel we show the contours of $r_u$ and $r_d$ in the $(m_A, \tan\beta)$ plane (red and blue lines, respectively). The orange, yellow and green regions refer to $\Delta <5$, $10$ and $20$ respectively, for $\Delta = {\rm max}(\Delta_{m^2_{\tilde t}}, \Delta_{X_t})$ (left panel) and $\Delta = \Delta_{m_A^2}$ (right panel). In the left panel, the white region is excluded by the request of positive stop-loop contributions and charge/color conservation ($|A_t|^2 \lesssim 3 (m_{Q_3}^2 + m_{t_R}^2)$~\cite{Casas:1995pd}). We do not show $r_V$ in our plots, since $|r_V-1| \ll |r_f-1|$ in most of the parameter space.

Some comments are now in order. Let us start from the consequences of the Higgs coupling measurements on the stop spectrum. 
The current experimental data from LHC8 still allow for ${\cal O}(10-15\%)$ deviations in $r_{GG}$ and in $r_{\gamma \gamma}$~\cite{CMS:2014ega, ATLAS}. From Fig.~\ref{fig:HiggsC}  we can estimate that the Higgs coupling measurements already require $m_{\tilde t} \gtrsim 300$ GeV depending on the mixing (see \cite{D'Agnolo:2012mj, Gupta:2012fy, Farina:2013ssa, Fan:2014txa} for more precise estimates), and it is thus already probing a certain part of the most natural region.   
This  bound  is more robust than those obtained from direct searches, since it does not depend on the details of the stop decay modes. 
On the other hand, assuming  that no relevant deviation will be observed,  LHC 13 will not significantly improve the current  bounds  on stops even with the $3000$ fb$^{-1}$ luminosity upgrade~\cite{Fan:2014txa}.  
At future machines such as ILC or TLEP, Higgs precision physics will be able to probe the couplings to gluons and photons up to a precision of about $1\%$ and $4\%$, respectively~\cite{1310.8361}. This means that, as shown in Fig.~\ref{fig:HiggsC}, future colliders will probe fine tuning regions up to $10\%$, since even deviations from the SM value as low as $1\permil$ are still compatible in some regions with a $10\%$ tuning.~\footnote{We explicitly checked that in this region there is no additional source in tuning from the Higgs couplings.}

If we cannot use precise Higgs measurements to rule out naturalness, we can vice versa use them to discover it: indeed a sizable deviation in the gluon gluon coupling can be accommodated only within the most  natural region, $ \Delta <5$.
A sizable deviation discovered in the next LHC run would need to be interpreted within this framework as the indirect sign of hidden light stops.

Let us now turn to the tree level couplings. From Fig.~\ref{fig:HiggsC}, right panel, we see that the coupling to the down-type quarks is the most constraining one, and already requires $m_A \gtrsim 400$ GeV, considering a current precision in the $ b b $ and $ \tau \tau $ couplings of about $(10-15)\%$.
However, we see from Fig.~\ref{fig:HiggsC} that for $\tan\beta \gtrsim 10$ deviations smaller than $ 0.5 \%$ are still compatible with $\Delta <5 $. 
This is true also in the MSSM, where the tuning is indeed set by the requirement of a $126$ GeV Higgs and not from requiring it to be SM-like. As with the gluon and photon couplings, precision measurements of the Higgs couplings are not powerful tools to rule out naturalness. They can however be used as probes of the model parameter space, giving bounds which are competitive with those that can be obtained from direct searches. More precisely, the HL-LHC will probe $r_d$ with a precision of about $4-7 \%$, while the ILC will be able to improve the precision up to $2\%$~\cite{1310.8361,McGarrie:2014xxa}. From the right panel of Fig.~\ref{fig:HiggsC} we see that such indirect searches at the HL-LHC will put a lower bound $m_A \gtrsim 800\, {\rm GeV}$, while the ILC will probe the multi TeV region. \\

Although up to now we have taken a bottom-up approach, we can argue on the challenges that a natural SM-like Higgs would imply for possible UV completions of this scenario.
A persistent agreement with the SM  predictions in the gluon gluon and $\gamma \gamma$ couplings can be accommodated as long as a sizable mixing $X_t$ is present. This would disfavor models where SUSY breaking is mediated primarily by gauge interactions.

Turning to the couplings with the SM fermions, the decoupling limit can be obtained in a natural way only for $|m^2_{H_u}| \ll |m^2_{H_d}|$, since we see from Eqs.~(\ref{eq:FT_general})-(\ref{eq:FT_general_largetb}) that even for large $\tan\beta$ an increase in $m^2_{H_u}$ has a fine tuning cost. For instance, $ m_{H_u} \sim 400 $ GeV already implies $\Delta_{m^2_{H_u}} \sim 20$. While from the bottom-up approach this is not too problematic, from the top-down we already see that in models which predict $m^2_{H_u}\sim m^2_{H_d}$ the decoupling limit can be challenging. An interesting direction to obtain the required hierarchy is given in~\cite{Csaki:2008sr},  where it is shown that the $m_{H_u}^2 \sim \mu^2 \ll B_{\mu} \ll m_{H_d}^2$ region is a possible natural solution to the $B_{\mu}-\mu$ problem in gauge mediation.
Another appealing possibility is to charge $H_d$, as well as the SUSY breaking mediators, under $SU(2)_B$ only. This model leads also to suppressed deviations of the Higgs couplings~\cite{McGarrie:2014xxa}. In this reference it is shown that in the chiral model the power of Higgs couplings measurement would be less effective, relaxing by $50-100$ GeV the possible reach both at LHC13 and at the ILC.

\section{Conclusions}

After the first run of the LHC, the naturalness of the electroweak scale is still under scrutiny. In particular, supersymmetric models must face the challenge both to meet the direct searches constraints and to accommodate a $126$ GeV SM-like Higgs boson. The two most studied supersymmetric extensions of the SM have different problematics, in this respect: in the MSSM the Higgs boson is naturally SM-like, but its mass requires a considerable tuning; on the contrary, in the NMSSM its mass is natural, but being SM-like may require some tuning. On top of this, although more model dependent, direct searches place lower bounds on the sparticle masses that must be taken into account when assessing the overall tuning of the theory. In this paper we only focused on the naturalness implications of the Higgs phenomenology, computing a lower bound on the overall tuning of the theory.\\

We have considered a supersymmetric scenario where the gauge sector of the MSSM is enlarged and the Higgs boson mass is increased at tree level via non decoupling $D$-terms. We focused as example on a simple non-abelian extension, $SU(2)_A \times SU(2)_B \times U(1)_Y,$  but our conclusions apply more broadly. 
We studied the fine tuning cost required to have a $126$ GeV SM-like Higgs, identifying and analyzing two sources of sensitivity:
the usual tuning on the electroweak scale, and the one on the scale at which the extended gauge sector is broken. The latter source put constraints on the parameters entering in the increased tree level Higgs quartic coupling, and it is therefore important to properly asses the fine tuning cost of raising the Higgs mass in this scenario.

From our analysis we can extract interesting conclusions. First of all, a $126$ GeV Higgs boson mass can be accommodated with an overall tuning better than $20\%$ for $\tan\beta \gtrsim 4$. This has to be compared with the MSSM, in which the main source of tuning is given by the stop masses needed to raise the Higgs boson mass up to the experimentally observed value. Moreover, although deviations are expected both in loop and tree level couplings, naturalness does not necessarily predict them to be large. In particular, we can compare the case under study with another natural extension of the MSSM, the NMSSM. The main difference is given by the $\tan\beta$ values needed to increase the Higgs boson mass at tree level in a natural way: $\tan\beta \gtrsim 4$ for the DMSSM, $\tan\beta \lesssim 3-4$ in the NMSSM. As we have seen, this implies that while Higgs precision measurements and heavy Higgs searches are powerful probes of naturalness in the NMSSM, for the DMSSM there can be heavy scalars without an effective fine tuning cost. 
A similar conclusion applies as well to SUSY models extended with triplets with hypercharge $Y = \pm 1$~\cite{Espinosa:1991gr}, since the supersymmetric coupling $W \supset H_u T H_u$ generates a $\lambda_u$ coupling in Eq.~(\ref{eq:potential_2HDM_loop}) whose contribution is maximized in the large $\tan\beta$ regime.

What are then going to be the naturalness probes in DMSSM models? In addition to higgsinos, stops and gluino direct searches, heavy gauge bosons are predicted by naturalness to have masses $m_{W'} \lesssim 6$ TeV, and to interact with matter with a coupling $g_{W'} \simeq g$ in the interesting region of parameter space. We defer to a future work a detailed analysis of the signals of such heavy bosons and of the expected LHC reach in such scenario.

\acknowledgments

 We would like to thank Raffaele Tito D'Agnolo, Marco Farina, Roni Harnik, Zhen Liu and Carlos Wagner for useful discussions, as well as the Mainz Institute for Theoretical Physics (MITP) for its hospitality and support. 

The work of E.B. is supported by the Spanish Ministry MICINN under contract FPA2010-17747.
Fermilab is operated by Fermi Research Alliance, LLC  under Contract No. DE-AC02-07CH11359 with the United States Department of Energy.

\appendix

\section{Electroweak precision measurements}\label{sec:appendix_EWPM}

Let us discuss here in some detail our analysis of EWPM. To this end, we notice that after the singlet $\sigma$ acquires its vev, $\langle \sigma \rangle = u$, but before EWSB, we can define massless ($W_\mu^a$) and massive ($Y_\mu^a$) gauge boson combinations as follows:
\begin{equation}
 \begin{array}{rcl}
  W_\mu^a &=& \frac{1}{\sqrt{g_A^2 + g_B^2}} \left( g_B A_\mu^a + g_A B_\mu^a \right)\\
  Y_\mu^a &=& \frac{1}{\sqrt{g_A^2 + g_B^2}} \left( g_A A_\mu^a - g_B B_\mu^a \right) \, ,
 \end{array}
\end{equation}
where $A_\mu^a$ and $B_\mu^a$  are the gauge bosons associated with $SU(2)_A$ and $SU(2)_B$, respectively. Apart from small corrections of order $v^2/u^2$, the heavy gauge bosons can be identified with the $(W^{'\pm}_\mu, Z_\mu')$ triplet.

The relevant terms in the massive gauge boson lagrangian can be written as 
\begin{equation}
 {\cal L} = -\frac{1}{2} m_{W'}^2 W^{'a}_\mu W^{'a}_\mu + g_{W'} W^{'a}_\mu J^a_\mu \, ,
\end{equation}
where $g_{W'} = g \frac{g_A}{g_B} $ and $m_{W'}^2= \frac{g_A^2 + g_B^2}{2} u^2$. The triplet current coupled to the massive gauge bosons is given by 
\begin{equation}
 J_\mu^a =   \overline{\ell}_L \gamma_\mu \tau^a \ell_L + \overline{q}_L \gamma_\mu \tau^a q_L +  i h^\dagger \tau^a \overleftrightarrow { D_\mu} h +  i H^\dagger \tau^a \overleftrightarrow { D_\mu} H + \dots \, ,
\end{equation}
where the dots represent the sparticle contributions, $h$ is the combination of scalar doublets acquiring vev $v$ and $H$ is the orthogonal combination.

Once we integrate out $Y^a_\mu$, the effective lagrangian is given by
\begin{equation}\label{eq:eff_lagrangian}
 {\cal L}_{eff} = \frac{g_{W'}^2}{2 m_{W'}^2} J_\mu^a J_\mu^a = \frac{g_{W'}^2}{2 g^2 m_{W'}^2} (D^\mu W^a_{\mu\nu})^2 \, ,
\end{equation}
where in the second equation we have used the $W^a_\mu$ equation of motion: $D^\mu W^a_{\mu\nu} = g J^a_\mu$. 
As is well known~\cite{Barbieri:2004qk,Falkowski:2013dza}, this operator generates only the $W$ parameter, constrained by LEP-II data. Using the numerical results given in~\cite{Barbieri:2004qk,Falkowski:2013dza}, we obtain the lower bound at $95 \%$ C.L.
\begin{equation}
 \frac{m_{W'}}{g_{W'}} \gtrsim 3.4\, {\rm TeV} \, ,
\end{equation}
which corresponds to the dashed line in Fig.~\ref{fig:gauge_masses}.


\begin{thebibliography}{99}
 
\bibitem{Barbieri:2000gf}
  R.~Barbieri and A.~Strumia,
  hep-ph/0007265.



\bibitem{Dine:2007xi}
  M.~Dine, N.~Seiberg and S.~Thomas,
  Phys.\ Rev.\ D {\bf 76} (2007) 095004
  [arXiv:0707.0005 [hep-ph]].



\bibitem{Hall:2011aa}
  L.~J.~Hall, D.~Pinner and J.~T.~Ruderman,
  JHEP {\bf 1204} (2012) 131
  [arXiv:1112.2703 [hep-ph]].



\bibitem{Fayet:1974pd}
  P.~Fayet,
  Nucl.\ Phys.\ B {\bf 90} (1975) 104.



\bibitem{Ellwanger:2009dp}
  U.~Ellwanger, C.~Hugonie and A.~M.~Teixeira,
  Phys.\ Rept.\  {\bf 496} (2010) 1
  [arXiv:0910.1785 [hep-ph]].



\bibitem{Espinosa:1991wt}
  J.~R.~Espinosa and M.~Quiros,
  Nucl.\ Phys.\ B {\bf 384} (1992) 113.



\bibitem{Espinosa:1991gr}
  J.~R.~Espinosa and M.~Quiros,
  Phys.\ Lett.\ B {\bf 279} (1992) 92.



\bibitem{Batra:2003nj}
  P.~Batra, A.~Delgado, D.~E.~Kaplan and T.~M.~P.~Tait,
  JHEP {\bf 0402} (2004) 043
  [hep-ph/0309149].



\bibitem{Bertuzzo:2014bwa}
  E.~Bertuzzo, C.~Frugiuele, T.~Gregoire and E.~Ponton,
  JHEP {\bf 1504} (2015) 089
  [arXiv:1402.5432 [hep-ph]].



\bibitem{Diessner:2014ksa}
  P.~Dießner, J.~Kalinowski, W.~Kotlarski and D.~Stöckinger,
  JHEP {\bf 1412} (2014) 124
  [arXiv:1410.4791 [hep-ph]].



\bibitem{Aad:2014kra}
  G.~Aad {\it et al.} [ATLAS Collaboration],
  JHEP {\bf 1411} (2014) 118
  [arXiv:1407.0583 [hep-ex]].



\bibitem{Chatrchyan:2013gia}
  S.~Chatrchyan {\it et al.} [CMS Collaboration],
  Phys.\ Lett.\ B {\bf 730} (2014) 193
  [arXiv:1311.1799 [hep-ex]].



\bibitem{TheATLAScollaboration:2013xia}
  The ATLAS collaboration [ATLAS Collaboration],
  ATLAS-CONF-2013-091, ATLAS-COM-CONF-2013-081.



\bibitem{Kribs:2007ac}
  G.~D.~Kribs, E.~Poppitz and N.~Weiner,
  Phys.\ Rev.\ D {\bf 78} (2008) 055010
  [arXiv:0712.2039 [hep-ph]].



\bibitem{Katz:2014mba}
  A.~Katz, M.~Reece and A.~Sajjad,
  JHEP {\bf 1410} (2014) 102
  [arXiv:1406.1172 [hep-ph]].



\bibitem{D'Agnolo:2012mj}
  R.~T.~D'Agnolo, E.~Kuflik and M.~Zanetti,
  JHEP {\bf 1303} (2013) 043
  [arXiv:1212.1165].



\bibitem{Gupta:2012fy}
  R.~S.~Gupta, M.~Montull and F.~Riva,
  JHEP {\bf 1304} (2013) 132
  [arXiv:1212.5240 [hep-ph]].



\bibitem{Farina:2013ssa}
  M.~Farina, M.~Perelstein and N.~Rey-Le Lorier,
  Phys.\ Rev.\ D {\bf 90} (2014) 1,  015014
  [arXiv:1305.6068 [hep-ph]].



\bibitem{Fan:2014txa}
  J.~Fan and M.~Reece,
  JHEP {\bf 1406} (2014) 031
  [arXiv:1401.7671 [hep-ph]].



\bibitem{Farina:2013fsa}
  M.~Farina, M.~Perelstein and B.~Shakya,
  JHEP {\bf 1404} (2014) 108
  [arXiv:1310.0459 [hep-ph]].



\bibitem{Gherghetta:2014xea}
  T.~Gherghetta, B.~von Harling, A.~D.~Medina and M.~A.~Schmidt,
  JHEP {\bf 1404} (2014) 180
  [arXiv:1401.8291 [hep-ph]].



\bibitem{Burdman:2014zta}
  G.~Burdman, Z.~Chacko, R.~Harnik, L.~de Lima and C.~B.~Verhaaren,
  Phys.\ Rev.\ D {\bf 91} (2015) 5,  055007
  [arXiv:1411.3310 [hep-ph]].



\bibitem{Craig:2013fga}
  N.~Craig and K.~Howe,
  JHEP {\bf 1403} (2014) 140
  [arXiv:1312.1341 [hep-ph]].



\bibitem{Batra:2004vc}
  P.~Batra, A.~Delgado, D.~E.~Kaplan and T.~M.~P.~Tait,
  JHEP {\bf 0406} (2004) 032
  [hep-ph/0404251].



\bibitem{Maloney:2004rc}
  A.~Maloney, A.~Pierce and J.~G.~Wacker,
  JHEP {\bf 0606} (2006) 034
  [hep-ph/0409127].



\bibitem{Medina:2009ey}
  A.~D.~Medina, N.~R.~Shah and C.~E.~M.~Wagner,
  Phys.\ Rev.\ D {\bf 80} (2009) 015001
  [arXiv:0904.1625 [hep-ph]].



\bibitem{Cheung:2012zq}
  C.~Cheung and H.~L.~Roberts,
  JHEP {\bf 1312} (2013) 018
  [arXiv:1207.0234 [hep-ph]].



\bibitem{Craig:2012bs}
  N.~Craig and A.~Katz,
  JHEP {\bf 1305} (2013) 015
  [arXiv:1212.2635 [hep-ph]].



\bibitem{Huo:2012tw}
  R.~Huo, G.~Lee, A.~M.~Thalapillil and C.~E.~M.~Wagner,
  Phys.\ Rev.\ D {\bf 87} (2013) 5,  055011
  [arXiv:1212.0560 [hep-ph]].



\bibitem{Athron:2009bs}
  P.~Athron, S.~F.~King, D.~J.~Miller, S.~Moretti and R.~Nevzorov,
  Phys.\ Rev.\ D {\bf 80} (2009) 035009
  [arXiv:0904.2169 [hep-ph]].



\bibitem{Athron:2012sq}
  P.~Athron, S.~F.~King, D.~J.~Miller, S.~Moretti and R.~Nevzorov,
  Phys.\ Rev.\ D {\bf 86} (2012) 095003
  [arXiv:1206.5028 [hep-ph]].



\bibitem{Athron:2013ipa}
  P.~Athron, M.~Binjonaid and S.~F.~King,
  Phys.\ Rev.\ D {\bf 87} (2013) 11,  115023
  [arXiv:1302.5291 [hep-ph]].



\bibitem{Bharucha:2013ela}
  A.~Bharucha, A.~Goudelis and M.~McGarrie,
  Eur.\ Phys.\ J.\ C {\bf 74} (2014) 2858
  [arXiv:1310.4500 [hep-ph]].



\bibitem{Lodone:2010kt}
  P.~Lodone,
  JHEP {\bf 1005} (2010) 068
  [arXiv:1004.1271 [hep-ph]].



\bibitem{Blum:2012ii}
  K.~Blum, R.~T.~D'Agnolo and J.~Fan,
  JHEP {\bf 1301} (2013) 057
  [arXiv:1206.5303 [hep-ph]].



\bibitem{Barbieri:1987fn}
  R.~Barbieri and G.~F.~Giudice,
  Nucl.\ Phys.\ B {\bf 306} (1988) 63.



\bibitem{Dimopoulos:1995mi}
  S.~Dimopoulos and G.~F.~Giudice,
  Phys.\ Lett.\ B {\bf 357} (1995) 573
  [hep-ph/9507282].



\bibitem{Kitano:2005wc}
  R.~Kitano and Y.~Nomura,
  Phys.\ Lett.\ B {\bf 631} (2005) 58
  [hep-ph/0509039].



\bibitem{Cassel:2009ps}
  S.~Cassel, D.~M.~Ghilencea and G.~G.~Ross,
  Nucl.\ Phys.\ B {\bf 825} (2010) 203
  [arXiv:0903.1115 [hep-ph]].



\bibitem{Cassel:2010px}
  S.~Cassel, D.~M.~Ghilencea and G.~G.~Ross,
  Nucl.\ Phys.\ B {\bf 835} (2010) 110
  [arXiv:1001.3884 [hep-ph]].



\bibitem{Aad:2014wea}
  G.~Aad {\it et al.} [ATLAS Collaboration],
  JHEP {\bf 1409} (2014) 176
  [arXiv:1405.7875 [hep-ex]].



\bibitem{CMS:2013bea}
  CMS Collaboration [CMS Collaboration],
  CMS-PAS-EXO-12-049.



\bibitem{Dimopoulos:2014aua}
  S.~Dimopoulos, K.~Howe and J.~March-Russell,
  Phys.\ Rev.\ Lett.\  {\bf 113} (2014) 111802
  [arXiv:1404.7554 [hep-ph]].



\bibitem{Godfrey:2013eta}
  S.~Godfrey and T.~Martin,
  arXiv:1309.1688 [hep-ph].



\bibitem{Gherghetta:2012gb}
  T.~Gherghetta, B.~von Harling, A.~D.~Medina and M.~A.~Schmidt,
  JHEP {\bf 1302} (2013) 032
  [arXiv:1212.5243 [hep-ph]].



\bibitem{Casas:1995pd}
  J.~A.~Casas, A.~Lleyda and C.~Munoz,
  Nucl.\ Phys.\ B {\bf 471} (1996) 3
  [hep-ph/9507294].



\bibitem{CMS:2014ega}
  CMS Collaboration [CMS Collaboration],
and studies of the compatibility of its couplings with the standard model,''
  CMS-PAS-HIG-14-009.

\bibitem{ATLAS}
  The ATLAS collaboration,
  ATLAS-CONF-2014-009, ATLAS-COM-CONF-2014-013.

\bibitem{1310.8361}
  S.~Dawson {\it et al.},
  arXiv:1310.8361 [hep-ex].



\bibitem{McGarrie:2014xxa}
  M.~McGarrie, G.~Moortgat-Pick and S.~Porto,
  Eur.\ Phys.\ J.\ C {\bf 75} (2015) 4,  150
  [arXiv:1411.2040 [hep-ph]].



\bibitem{Csaki:2008sr}
  C.~Csaki, A.~Falkowski, Y.~Nomura and T.~Volansky,
  Phys.\ Rev.\ Lett.\  {\bf 102} (2009) 111801
  [arXiv:0809.4492 [hep-ph]].



\bibitem{Barbieri:2004qk}
  R.~Barbieri, A.~Pomarol, R.~Rattazzi and A.~Strumia,
  Nucl.\ Phys.\ B {\bf 703} (2004) 127
  [hep-ph/0405040].



\bibitem{Falkowski:2013dza}
  A.~Falkowski, F.~Riva and A.~Urbano,
  JHEP {\bf 1311} (2013) 111
  [arXiv:1303.1812 [hep-ph]].




\end{thebibliography}
\end{document}